# Large-scale Array for Radio Astronomy on the Farside (LARAF)

Xuelei Chen[1,4,7,8,*], Feng Gao[2,*], Fengquan Wu[1,8], Yechi Zhang[2], Tong Wang[2], Weilin Liu[2], Dali Zou[2], Furen Deng[1,7], Yang Gong[1], Kai He[1,8], Jixia Li[1,8], Shijie Sun[1,8], Nanben Suo[3], Yougang Wang[1,8], Pengju Wu[4], Jiaqin Xu[1,8], Yidong Xu[1,8], Bin Yue[1,8], Cong Zhang[1,7], Jia Zhou[5], Minquan Zhou[6], Chenguang Zhu[1,7], Jiacong Zhu[1,7]

1. National Astronomical Observatories, Chinese Academy of Sciences, Beijing 100101, China
2. Beijing Institute of Spacecraft System Engineering, Beijing100094, China
3. Tibet University, 10 Zangda East Road, Lhasa, Tibet 850000, China
4. Department of Physics, College of Sciences, Northeastern University, Shenyang, 110819, China
5. Shanxi University, Taiyuan, Shanxi 237016, China
6. Hangzhou Dianzi University, Hangzhou, Zhejiang 310018, China
7. University of Chinese Academy of Sciences, Beijing 100049, China
8. Key Laboratory of Radio Astronomy and Technology, CAS, Beijing 100101, China

# Summary

At the Royal Society meeting in 2023, we have mainly presented our lunar orbit array concept called DSL, and also briefly introduced a concept of a lunar surface array, LARAF. As the DSL concept had been presented before, in this article we introduce the LARAF. We propose to build an array in the far side of the Moon, with a master station which handles the data collection and processing, and 20 stations with maximum baseline of 10 km. Each station consists 12 membrane antenna units, and the stations are connected to the master station by power line and optical fiber. The array will make interferometric observation in the 0.1-50 MHz band during the lunar night, powered by regenerated fuel cells (RFCs). The whole array can be carried to the lunar surface with a heavy rocket mission, and deployed with a rover in 8 months. Such an array would be an important step in the long term development of lunar based ultralong wavelength radio astronomy. It has a sufficiently high sensitivity to observe many radio sources in the sky, though still short of the dark age fluctuations. We discuss the possible options in the power supply, data communication, deployment, etc.

*Author for correspondence (xuelei@bao.ac.cn, I.am.gaofeng@Gmail.com ).

## 1. Introduction

Astronomy is a fundamentally observational science, its advance is driven largely by the progress in observations. Throughout history, new discoveries in astronomy changed our view of the Universe, and gave us many deep insights on the fundamental laws of Nature. However, at present there is still a part of the electromagnetic spectrum for which the astronomical observation is very limited--this is the frequency band below 30 MHz, which we shall refer to as the ultralong waveband. Observation in this band is very difficult from the ground, due to the absorption, reflection and strong refraction of the low frequency radio wave by the ionosphere, and also because of the large amount of radio frequency interference (RFI). Thus, since 1960s, there had been proposals for making astronomical observations from the far side of the Moon [1-3]. With advances in space technology, it is now conceivable to take advantage of the ideal condition offered by the far side of the Moon, where the ionosphere is avoided and the RFI from the Earth can be well shielded, to open up a new observational window at the ultralong wave band.

At the Royal Society meeting Astronomy from the Moon: the next decades in 2023, we have mainly presented our lunar orbit array concept, called the Discovering Sky at the Longest wavelength (DSL). The basic concept of the DSL is a linear array of micro-satellites in a circular orbit around the Moon. These satellites will be launched as an assembly by a single rocket, and released in the orbit to form a linear array. A "mother" satellite at one end of the array will collect the observational data taken by the other "daughter" satellites when the array is on the part of orbit where the Earth is shielded by the moon, and relay the data to Earth once the array comes to the part of the orbit where the Earth is in view. The daughter satellites will take both the global spectrum of the sky to probe the cosmic dark ages and dawn, and also make interferometric imaging observation of the sky. A description of the DSL concept has been given in a paper[4] prepared for the 2020 meeting on *Astronomy from the Moon: the next decades*, and published in the earlier *philosophical transactions* collection on the same topic[5]. Some scientific and technological problems associated with the DSL are also studied in a number of other papers[6-10].

In the present paper, we will not repeat the description of the DSL concept. Instead, here we consider the basic concept of a lunar surface array, called the Large-scale Array for Radio Astronomy on the Farside (LARAF). The basic idea here is an interferometer array at the far side of the Moon, based on the current technology and engineering capability.

Engineeringly, deploying a radio telescope on the lunar surface is far more complicated than deploying one in the lunar orbit. Nevertheless, there are also some advantages of building a low frequency radio array on the lunar surface over the lunar orbit. The relative positions of the units are fixed, there is no need to worry about controlling its position at all times, and the surface offers a stable platform, which is especially important to observe variable sources. There have also been a number of recent proposals for lunar surface radio telescopes, such as the ROLSS[11], LRA/DALI[12], LuSEE-night[13], FARSIDE[14] ALO [15], and the Lunar Crater Radio Telescope (LCRT)[16]. Of course, there are also disadvantages, such as the charged dust near the lunar surface and the lunar regolith which may also affect the radio observation.

Below, we describe the basic science goals, then come up with a basic concept design, and discuss the various engineering considerations and technical choices.

## 2. Science Goals

The ultralong wave observation will open up a new window for astronomy, as there is very little data in this band at present, and the few observations are also severely limited by their angular resolution and sky coverage. Based on what we know to be likely sources of low frequency radio emission, the following areas are of possible interest: cosmic dark ages and cosmic dawn, quasar and radio galaxies, interstellar medium and cosmic ray, the Sun and planets, stars and exoplanets. These have been discussed extensively in the literature(e.g. [17]), and also reviewed in the papers cited above.

## 2.1 Dark Ages

A key science goal in the low frequency radio band is to observe the cosmic dark ages, i.e. the era between Recombination and the formation of first stars and galaxies. Here it can play a unique role, as it is the only direct observational probe of this era in the cosmic history. During the dark ages, the structure formation is still at its early stage, with minimum amount of non-linear evolution. With its large comoving volume and linearly evolved perturbations, the dark ages preserve huge amount of information about the origin of the Universe in the form of primordial fluctuations, which could be generated during the era of inflation[18]. Photons of 21cm wavelength are emitted or absorbed by the neutral hydrogen atoms against the cosmic microwave background (CMB), depending on whether the hydrogen spin temperature is higher or lower than the CMB temperature at the time, and those spectral features generated during the dark ages are now redshifted to the low frequency bands. The epoch of Recombination has a redshift of about 1100, corresponding to a redshifted frequency of 1.3 MHz, though in standard model we only expect significant 21cm signal below redshift $z \sim 200$ (frequency > 7MHz). The redshift of the end of dark age at which the first stars formed in large amount is much less certain, as it depends on the complicated astrophysics of star formation. We tentatively take it to be around redshift 30, corresponding to frequency 47 MHz.

The hydrogen spin system is coupled both to the CMB by the emission or absorption of the 21cm photons, and to the kinetic motion of the gas by either atomic collisions, or by the scattering of Lyman series photons, a.k.a. the Wouthuysen-Field mechanism [19,20] at the later epoch of cosmic dawn, when the first stars and galaxies formed. During the dark ages, the gas in the bulk is cooled by its expansion, but heated by its compression in the first formed clumps, and later on heated by the ionizing photons from the first stars and galaxies. The coupling between the spin and kinetic temperature also varies as the atomic density and the Lyman photon background varies. The 21cm signal is modulated by the ionization state, density and spin temperature of the hydrogen [21,22].

There are two modes of observation for the 21cm signal from dark ages: the 21cm global spectrum, and the 21cm fluctuations. In the global spectrum measurement, the sky-averaged 21cm brightness temperature is measured. This probes the evolution history of the dark ages. In many cases, we expect an absorption trough associated with the cosmic dawn, when the spin temperature is coupled by the Lyman photons from the first stars and galaxies to the low kinetic temperature of the gas[23,24], though the detailed spectrum depends on the cosmological model. This measurement can be carried out using a single antenna with precisely calibrated instrument response. The detection of a possible absorption trough has been claimed by the EDGES experiment]25], though the absorption strength exceeds the prediction of the standard model, and this result is in conflict with the measurement of the SARAS-3 experiment[26].

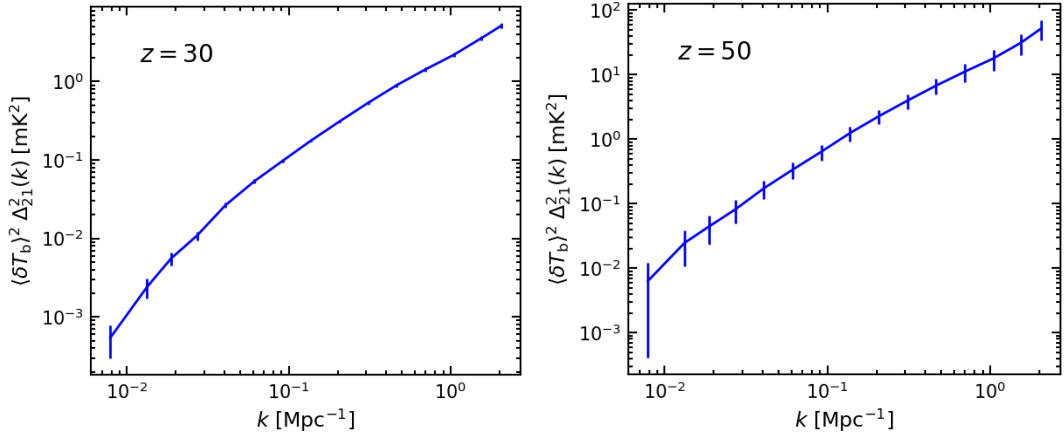

Figure 1.    The expected 21cm power spectrum at z=30 and 50, the error bars show the expected noise power for a 10km² array, 1MHz bandwidth and one year integration time.

The spatial fluctuations of the 21cm can be considered as the ultimate source of information... i ... angular resolution ... iation is the str... For ar... e power spectr...

$$P_{noise} = \frac{\lambda^{k}}{\pi} \frac{(D_c^2 l_z \Omega_{FOV}/N_{beam})}{\Delta \nu \, t_{int}} \frac{T_{sys}^2}{A_{coll}^2}$$

where $D_c$ the comoving distance, $l_z$ is the redshift range corresponding to the bandwidth, $\Omega_{FOV}$ = ... $T_{sys}$ the sys... e area for ea... y that the $uv$ ... ting areas ... re, the expect... egration time of one year. If the integration is carried out during the lunar night, it would require 2~3 years of observation of the array.

We see detecting the primordial fluctuations in the dark ages require a very large area to acquire sufficient sensitivity [20]. The development of such an array in the far side of the Moon would be a great feat in technology. Probably we need to approach such a grand goal in a few steps: first test the basic technology with a few dipole antennas ($10^2$ m² collecting area), then move on to an array of a few tens or a few hundred antenna ($10^3$~$10^5$ m²), and finally the huge antenna array with $10^3$~$10^4$ antennas (>$10^6$ m²) that is capable of observing dark ages fluctuations.

Below we consider the intermediate step: an array of $10^2$ antennas, which is short of dark age 21cm fluctuation measurement, but can already be considered as a large array, and can make many astronomical observations.

## 2.2 Astronomical Observations

An array of $10^2$ antennas can already observe a large number of radio sources in the ultralong wavelength band. We consider an array of dipoles. As the wavelength is very large, it is often difficult to construct antenna of optimal size. Furthermore, the relative bandwidth is also very large, so even if we have an antenna being optimal size at one wavelength, it is not so at other frequencies. If the dipole is electrically short, i.e. with size $L_a \ll \lambda$, the direction gain and effective aperture of such an antenna are

$$D = \frac{3}{2}, \qquad A_e = \frac{3}{8\pi}\lambda^2$$

8However, the radiation resistance of the antenna is small in such case, $R_{rad}=20\pi^2 (L_a/\lambda)^2$, so in the end, only a small fraction of the radiation power could be received, for the incoming flux S, the received power is

$$P = \eta\, A_e\, S$$

where $\eta=R_{rad}/(R_{rad}+R_{loss})$. While the exact value depends also on the circuit, generally we would have $\eta \sim \lambda^2$, so in the end if we denote the effective collecting area $A_c = \eta\, A_e$, then the received power $P \sim A_c \sim L_a^2$. Neglecting auto-correlation, the sensitivity of the interferometer array for an unpolarized point source can be written as

$$\sigma_S = \frac{k_B\, T_{sys}}{A_c \sqrt{2N(N-1)\,\Delta\upsilon\, t}}$$

If the system is well-matched, i.e. observing at the wavelength comparable to the physical size of the antenna, then this is reduced to the usual formula with $A_c=A_e$, but in the short antenna limit $A_c \sim L_a^2$. At the low frequency the system temperature is dominated by the sky temperature, for although P suffered a loss by the factor η as noted above, at low frequency where the sky is very bright the effective antenna temperature is still higher than the receiver thermal noise.

In Fig.2 we plot the sensitivity as a function of the number of antennas at 10, 30, 50 MHz for a bandwidth of 1 MHz, and an observation of 100 hours. Such an antenna array will be able to observe many interesting astrophysical objects. For example, from Fig.2, at 10 MHz the flux the limit is about 0.1 Jy for a $10^2$ antennas. Cygnus A has a flux of 13.5 kJy at 10 MHz [30], its distance is 162 Mpc/h, so with the above sensitivity we should be able to detect such a radio galaxy at cosmological distance. Most radio galaxies has lower radio luminosity than Cyg A, but may still be observable at large distances. The radio emission of the radio galaxies comes mainly from the synchrotron emission of energetic electrons in the jet from the central supermassive black hole. As the activity of the black hole ceases, the electrons cool down, and their radiation frequency lowers. It is expected that the ultralong

*Phil. Trans. R. Soc. A.*8

wavelength observation will enable us to probe the relics of these jets, at least for the nearby ones, and obtain a more complete view of the black hole activity in the Universe.

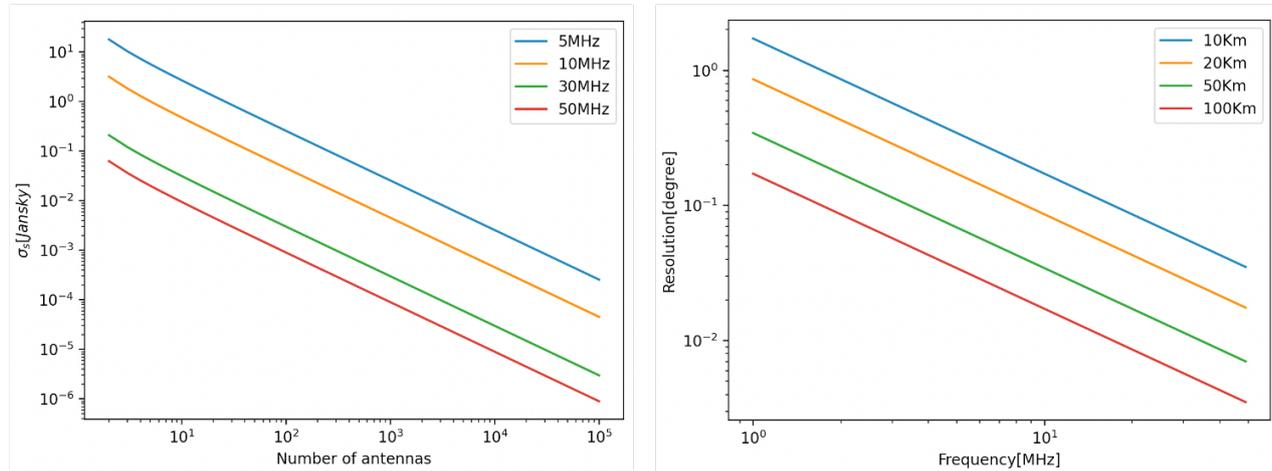

Figure 2.	Left: Sensitivity of an array of 10m antenna, as a function of antenna number, for bandwidth of 1MHz and integration of 100 hours. Right: Angular resolution as vs. frequency, for baseline of 10,20,50 and 100km.

Exoplanet is another interesting area where the radio array at the far side of the Moon can make new discoveries. In the radio band, the exoplanets with strong magnetosphere may produce low frequency radio emission with flux comparable to the host star, especially during the flares where the emission is greatly enhanced. Some nearby exoplanets should be detectable, given their expected peak flux [31,14]. However, so far the ground based low frequency observations have only found weak emissions, and in the case of HD 80606b, no emission is detected down to a level of a few mJy at 50 MHz[32], the space observation at lower frequencies would help shed light on this problem.

The angular resolution of the array is given by $\theta=\lambda/b_{max}$, where $b_{max}$ is the maximum baseline length. We plot the angular resolution as a function of frequency with $b_{max}$ =10, 20, 50 and 100km. At 10km and 30 MHz wavelength, the angular resolution is about $10^{-3}$ rad or 3.4 arcmin. This resolution is capable of resolving some nearby radio galaxies and supernovae remnants, but perhaps inadequate for the more distant ones. Increasing the baseline length would be useful. However, the maximum angular resolution is limited by the ISM and interplanetary medium (IPM) scattering. The angular broadening due to ISM and IPM scattering are [17]

$$\theta_{ISM} = \frac{30'}{(\nu/MHz)^{2.2}\sqrt{\sin b}}, \quad \theta_{IPM} \approx \frac{100'}{(\nu/MHz)^2}$$

Due to the influence of dispersion and scattering in the ISM and IPM, the time domain broadening of the time-varying signal in the ultra-long wave band are [17,33]

$$\text{ISM: } \Delta t = 6 \text{ yr } (\nu/MHz)^{-4.4}$$

$$\text{IPM: } \Delta t = 0.1 \text{ s } (\nu/MHz)^{-4.4}$$

It is not very feasible to measure time-varying signals for signals outside the solar system on time scales of interest. Therefore, there is no strict requirement for the observation time resolution.

## 3. Design Considerations

We consider an antenna array on the far side of the Moon. The array shall be composed of a number of sub-array or stations, each containing a number of antennas. Many ground-based low frequency arrays have adopted this form, e.g. the LOFAR, MWA and SKA-low. It allows widely distributed antennas, while concentrating the electronics at relatively few places. The signal at each station will be digitized, fast Fourier transformed, and filtered and packaged to send to the master station. A master station, which may coincide with the lander, will be responsible for collecting the data from all stations, correlate the voltage data to produce interferometric visibilities, pre-process and store the results, and transmit the data to the relay for downlinking to Earth.

Including both auto-correlations and cross correlations, there are $N(N+1)/2$ visibilities. If the integration time is T, the rate for transmitting the visibility data would be

$$R = \frac{N(N+1)N_f L}{2T}$$

where $N_f$ is the number of frequency channels, and L is the number of bits used for recording each visibility. The Moon has a very slow rotation rate ($3\times10^{-6}$ rad/s), for a baseline of ~10km, this allows an integration time of over $10^2$ s if we require the baseline swipe less than 1/10 of the wavelength. For example, if N=252, and $N_f$=2048, and we use 24 bits to transmit visibility for each frequency channel, the required data rate is about 15 Mbps. For some special observations, e.g. those of transient objects, or engineering tests, one may need to adopt a much shorter integration time, or even transmitting the raw data. In those cases such data rate is not enough, but as such observations will only happen occasionally, the data can be transmitted over a longer period.

To reduce the total mass to be shipped to the Moon, we consider the antenna constructed with the lightest material. At present, this will probably take the form of membrane antenna. Such an antenna could be deployed with a reel. The reel will be carried and deployed to the designated place by a lunar rover, which will unroll itself to form the station.

The stations could be connected to the master station by wired connection, or by wireless data link. While the wireless connection allows more freedom in the deployment site and route, the wired connection generally allows more bandwidth for data communication, and avoids generating radio frequency interference in the lunar far side. Furthermore, power may also be supplied from the master station using the same cable. The options of power supply will be discussed in more detail later. After weighing the various options, we decide to adopt the wired connection. The cables will be deployed from the master station by the rover en route to each station site. Care has to be taken to design the cable system to avoid generating RFI. To mitigate the interaction between the cable and antenna, we may also need to insert some radio frequency chokes along the cable.

## 3.1 The Antenna Stations

Each station has a total 12 antennas. To reduce the weight, membrane antenna is used, which can be packed as rolls, and unrolled from the station. The membrane will serve both as the base for the antennas, and also for power and signal connection to the station center. When deployed, these will form a subarray of antenna units. In Fig.3, we show two design options which we are studying. Each antenna has a length of 10 meters, made by conducting layer on the membrane.

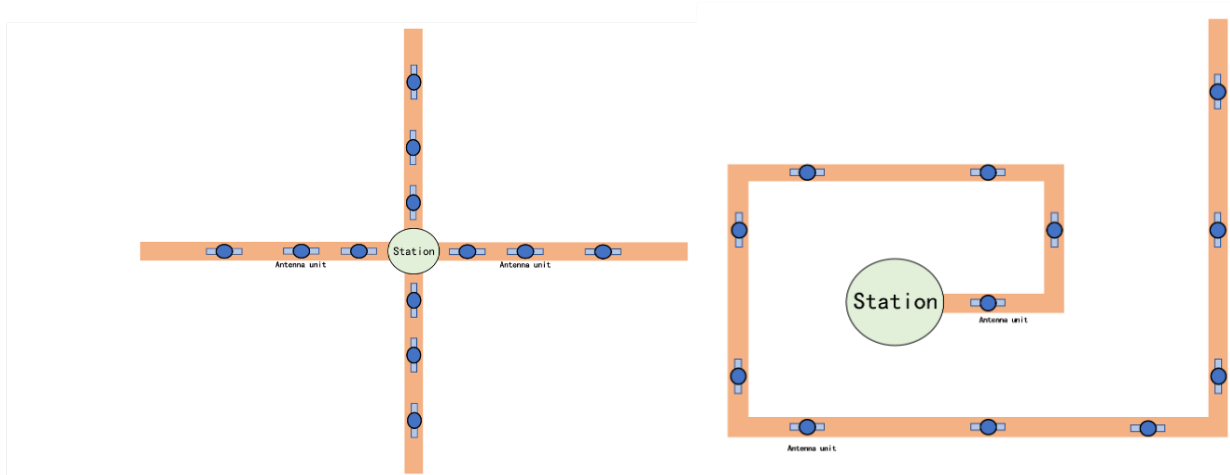

Figure 3.  Two designs of station antenna units on long membrane.

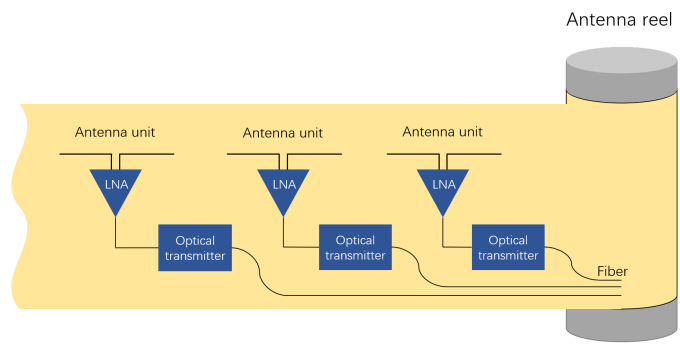

Figure 4.  Schematic diagram of antenna unit.

Each antenna unit has a receiving module on the membrane, including amplifiers, calibration unit, analog optical transmission unit, and thermal control module. The radio frequency (RF) signal will be converted to optical signal, and send to the station center via optical fiber, as shown by the scheme diagram in Fig.4. The outlook and internal structure of the receiving unit is shown in Fig.5. It is designed to work during the lunar night or the low temperature part of the lunar day. During the high temperature part of the day it will be powered off. It may be possible to shield the electronics a little by burying it slightly, which would reduce the temperature variation during the day-night cycle, but would require additional capability for the rover during the deployment.

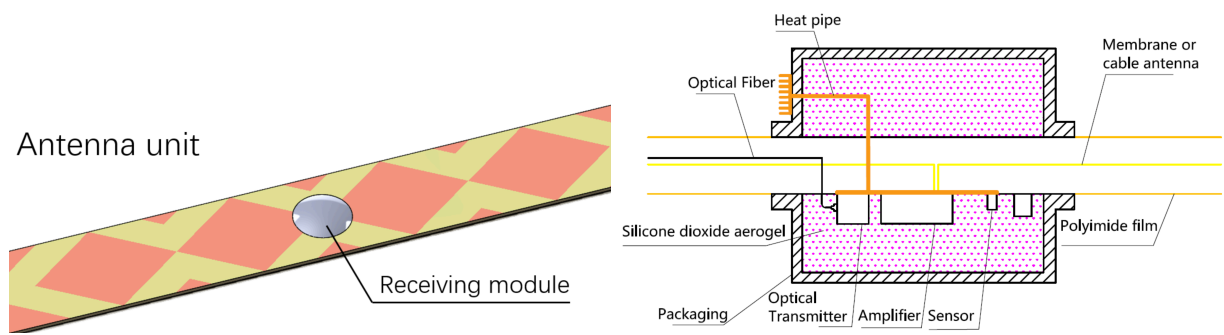

Figure 5.     The receiving module of antenna unit (Left) and its internal structure (right)

At the station center the signal is digitized, and pre-processed, as shown in Fig.6. The station also has to ensure the synchronization of the clocks.  The master station, also shown in Fig.6,  besides having the functions of other stations, can also send command to the other stations, and it can collect the data from other stations, and make correlations to generate the interferometric visibility data, or form digital beams. It should also be able to store the data, and forward this to the relay satellite.

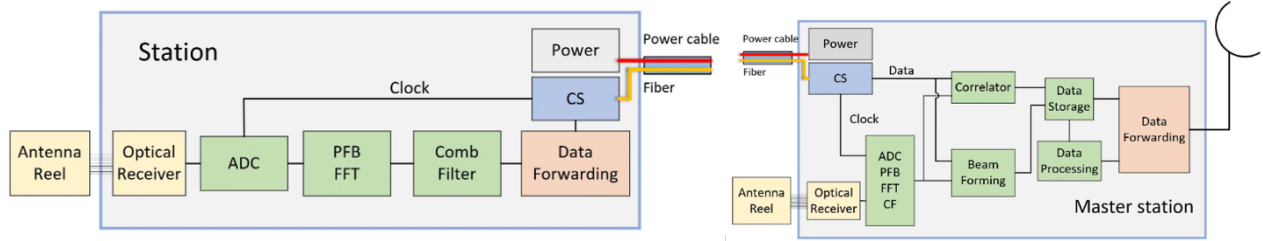

Figure 6.     Block diagram of the master station (Right) and a station (Left)

## 3.2 Array parameter and configuration

Based on the science objectives, we set the working frequency range as 0.1-50MHz tentatively. For probing the dark ages, perhaps the 20-50 MHz part is most interesting. However, as noted in Sec.2, at present it is difficult to realize an array with sufficient size for this purpose. On the other hand, for exploring the new observational window, the frequency below 10 MHz is most interesting, as this is the part of the spectrum least explored.

From the perspective of project planning and carrying capacity development, an array of about 252 antenna elements with a total load weight of about 1.5 tons could be realized in the near term (2030). The main parameters are as follows:

- Working Frequency Range: 0.1-50 MHz
- Total load weight: 1-1.5 tons
- antenna length: 10 m
- 1 master station, 20 stations，each with 12 antenna units, a total of 252 antenna units.
- Maximum Baseline: 10 km
- Weight of master station 400kg
- Weight of stations: 20kg /station
- Lunar Rover: 120kg

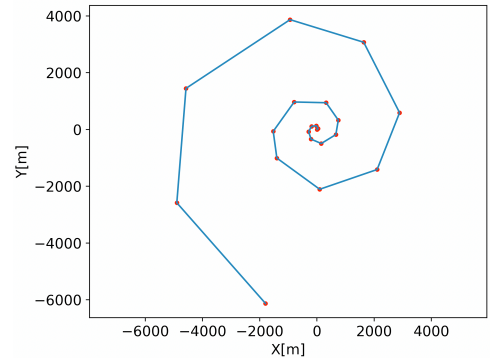

Figure 7.    Station Distribution of the Array

The array is planned to have a maximum baseline of 10 km, and the array configuration is optimized to make the main lobe sharp and sidelobes suppressed by having uv coverage on different scales. A possible array configuration (spiral) is shown in Fig.7. The stations form a spiral distribution to achieve a Gaussian distribution in the UV plane (Fig.8 Left). In this way, the sidelobes could be suppressed, as shown in the right panels of Figure 8.

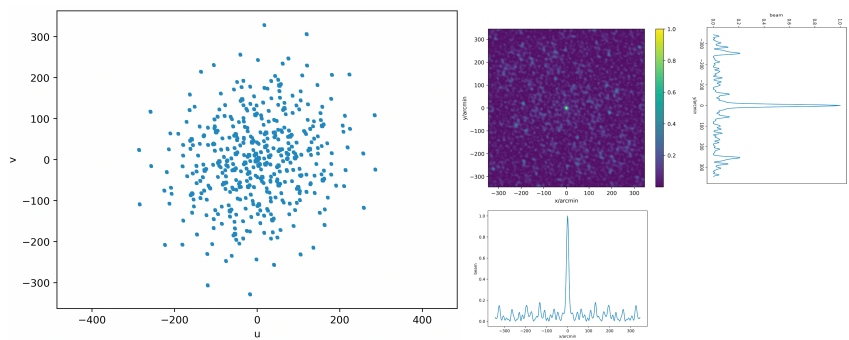

Figure 8.    Left: antenna UV coverage, right: antenna two-dimensional beam pattern and projection of two directions

## 3.3 Power supply

We have considered the arrangement of the working time for the array. Except for the observation of the Sun, nighttime observation is generally more desirable, as the active Sun can affect observations when it is above the horizon. Furthermore, for the instruments which are laid in the open, heat dissipation is also a great challenge if we are to observe during the lunar day, which has a very high temperature. On the other hand, the very low environment temperature during lunar night does not pose a serious problem. We therefore set the working time of the array to include the whole lunar night, and part of the day (⩽25℃), which is about 62% of a lunar day. In Figure 9 we show the evolution of the lunar surface temperature as measured by the Chang'e-4 lander at 177.6°E and 45.5°S [34], and the operating time for the above condition is marked by green color, though of course for a different site there will be some differences. During the rest of the lunar day time (⩾25℃), the array enters the standby mode, and only the data transfer, TT & C, computation and thermal control of the master station consume energy. This is different from the conventional lunar rover mode, which works during the lunar day and sleeps during the lunar night.

The continuous full-power operation in the 14.5 days of no-sunlight environment and the extremely low temperature of -190℃ lead to a very large demand for electric energy.

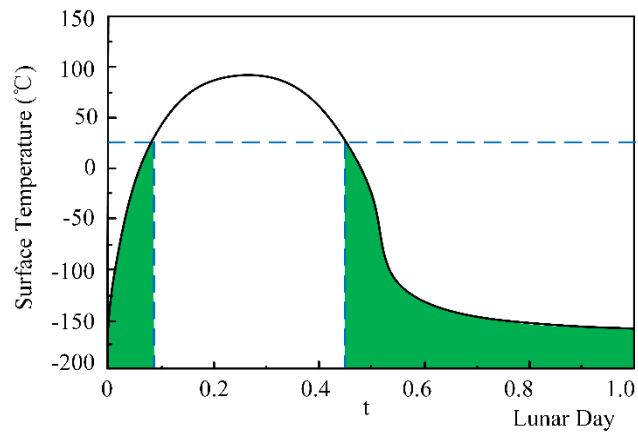

Figure 9.    The CE-4 measured lunar surface temperature and array operating time (green).

Through careful design, and application of ultra-low power digital components, etc., the power consumption of a single antenna is reduced as much as possible, and high voltage is used to reduce the voltage drop during the long-distance transmission. The total power budget of the master station, station and all antennas is 620W during the lunar night, but it still requires more than 220,000Wh of battery energy to get through the long lunar night.

The energy system generates, stores and supplies the energy. It includes:

- Orbiter energy subsystem: supplies the power for the digital communication of the orbiter. A larger capacity is required if wireless energy transmission is used.
- Lander (master station) energy subsystem: supplies the energy of the data transmission between the master station and the orbiter, data processing, and thermal insulation in the lunar night.
- Lunar rover energy subsystem: supplies the energy for the antenna deployment and thermal insulation during the lunar night;
- Radio array (station, antenna) energy sub-system: supplies the energy for observation and insulation.

There are two energy supply options: 1) Centralized energy supply: all energy is delivered from the central master station to each station, as shown in Fig. 10. This option is particularly suitable for some forms of power generation, such as the nuclear reactor power systems and the regenerated fuel cells (RFCs). 2) Distributed power supply: the master station and stations each have an independent power supply.

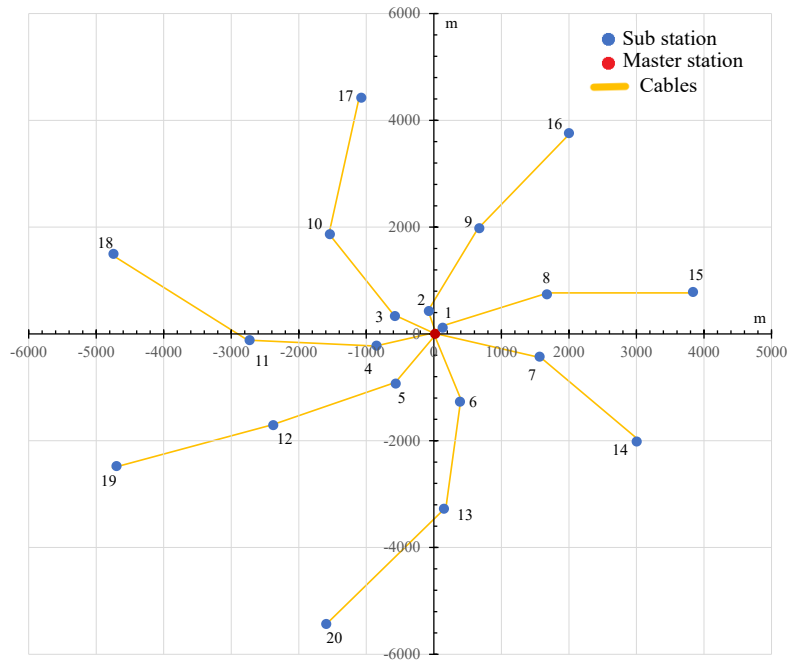

Figure 10.    Cable routing for centralized power supply between master station and stations

The options for energy generation include: solar panels, Li-ion batteries, regenerated fuel cells (RFCs), radioisotope thermo-electric generator (RTG), nuclear reactor power systems, laser wireless energy transfer (laser transmitter, laser receiver), power cables, etc. The orbiter, lunar rover, master station, stations and antennas can use different options, and we can use the permutations and combinations of these to form different technology routes. We make the following assumptions:

(1) The lunar rover will use the solar panels for power supply during the lunar day, and return to the master station for insulation during the lunar night. The master station uses fuel cells to provide energy during the lunar night, with an energy demand of about 150W.

(2) RFCs and nuclear reactor power systems are more suitable for centralized high-power supply from the master stations, not suitable for distributed power supply.

Besides the lunar rover which is powered by the solar panels, and the antennas, which are powered by cables connected to the station, there are 11 options for the rest of the spacecraft/equipment composition. The choice of the technology route is determined by considering the following three aspects: the lunar landing mass, technology maturity, and accessibility. The mass of each spacecraft for these 11 solutions were calculated, and the results are shown in Fig. 11.

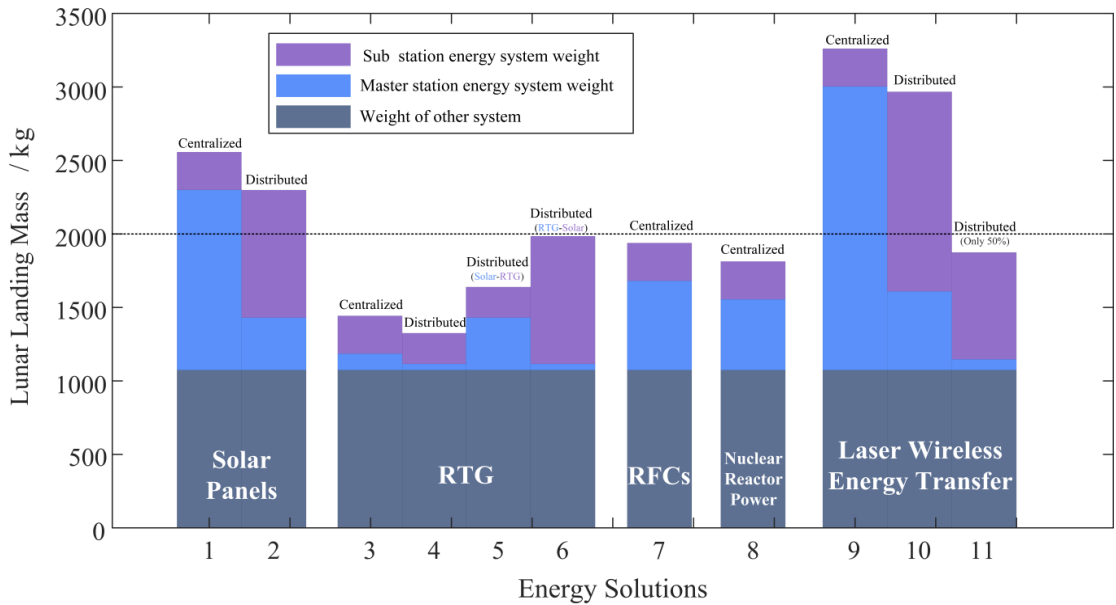

Figure 11.     Comparison results of energy solutions

From the above analysis, the RTG is the technical route with the lightest weight, and has a high TRL, but the Pu 238 are difficult to obtain. The distributed wireless energy transfer can reduce the use of cable, but due to the heavy weight of the equipment, it does not bring significant weight advantages, and its TRL is low.

The RFCs seems to be the optimal solution. It has a high TRL, and a higher energy density than the traditional Li-ion batteries, with a total weight of about 2000 kg for all lunar lander equipment. The RFCs in the master station will use solar energy to electrolyze water during the day, and generate electricity at night through a combustion reaction. The master station supplies power to the other stations through the cables shown in Fig. 10. There are seven cable cantilevers, each with an average length of approximately 6 km, and the fiber optic and power cables are integrated together, weighing about 4 kg/km.

## 3.4   Deployment

According to the overall design, the 20 stations will be deployed in a helical array configuration, as shown in Fig. 7. However, if these stations are deployed along the spiral line using a lunar rover in one pass, a high load-bearing capacity is required for the rover to carry the 20 stations. Also, unless the 20 stations could be deployed during one lunar day which would require very high speed, the rover would be required to stay out and survive the lunar night en route.

Therefore, a more viable option is adopting the radial path shown in Figure 10, where the spiral array is decomposed into seven segments, each containing two to three stations, and will be deployed in turn. After the lunar rover completes the deployment of antenna stations on each path, it will return along the path to the central station to reload the stations for the next one. The conceptual diagram of lunar rover deployment is shown in Fig.12. The position of each station relative to the center will be measured by the rover when it is deployed.

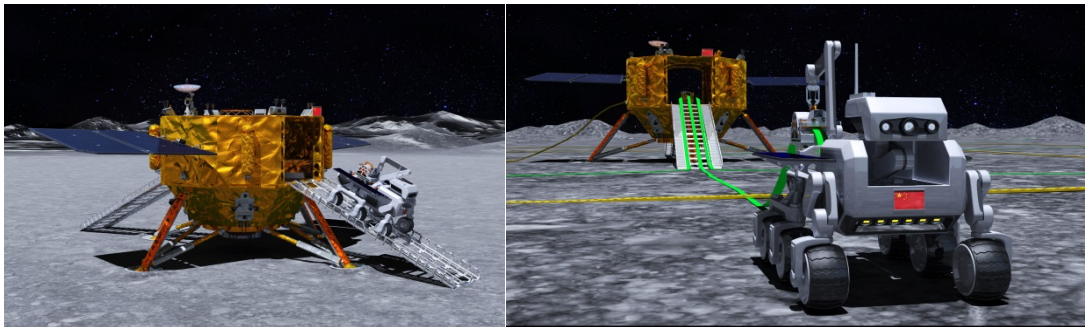

Figure 12.         Conceptual diagram of lunar rover deployment

The moveable distance of the rover is at least 15km per lunar day, and the maximum driving speed is not less than 40m/h. The lunar rover needs following functions:

(1) Autonomous path planning and obstacle avoidance, with a climbing capacity of not less than 20°; (2) High-precision relative position measurement; (3) High-precision visual image 3D inversion and measurement; (4) Station handling and release capability, with a carrying capacity of not less than 80kg; (5) Station fixing and antenna node deployment with dexterous manipulators; (6) When necessary, remote operation can be performed to assist station and antenna deployment.

To accomplish its task, the rover is designed with the following operating modes:

(1) Perception mode. The navigation camera is powered up to scan the surrounding and obtain navigation data, and processes it to generate the planned path.

(2) Mobile mode. Receive control commands, arrive at staged navigation point, and complete local path planning independently as needed.

(3) Autonomous driving mode. Move autonomously with the guidance, navigation and control (GNC) subsystem to achieve a single travel distance $\geqslant 8$ m, and an autonomous multiple continuous travel distance $\geqslant 100$ m.

(4) Operation mode. The robotic arm of the lunar rover is powered up to load, place, and unfold the antenna units.

(5) Charging mode. The rover is kept stationary, the solar wing is adjusted to achieve solar orientation, and the battery starts charging.

(6) Hibernation mode. The rover is powered off, except for waiting to receive commands.

## 3.5 Relative position determination

The GNC subsystem of the lunar rover could measure the relative position between the lunar rover and the master station. As shown in Figure 13, the GNC subsystem consists of a global planning sensor, two local planning sensors, a laser rangefinder, two star sensors, two inertial measurement sensors and a navigation control unit, to carry out the attitude determination, environment sensing, path planning, motion control, safety monitoring and fault diagnosis.

The accuracy of attitude measurement is better than 2° (3σ) during movement, and 1° (3σ) at rest. The relative distance measurement accuracy is better than 0.5m.

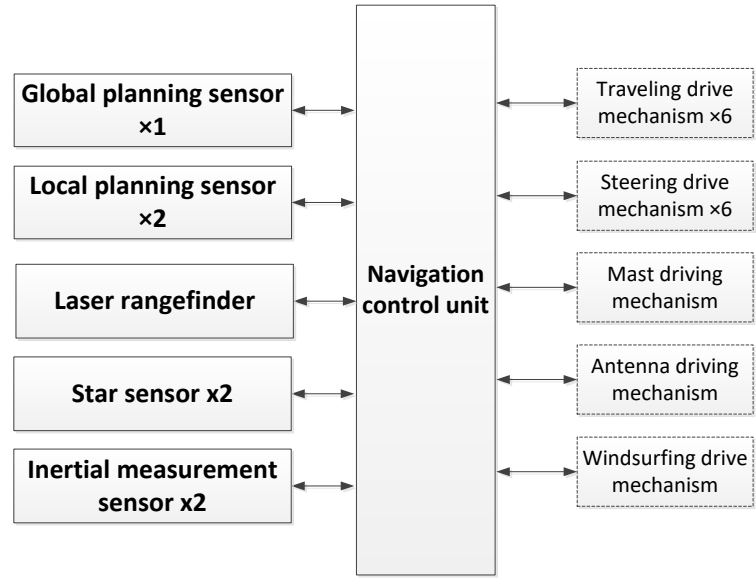

Figure 13.    GNC system components

The GNC subsystem achieves long-distance movement through autonomous path planning and local obstacle avoidance planning. Taking the final target point, the global path planning algorithm is performed at each stop, and generates a series of path points within the current sensing range. These path points are connected to form a fold trajectory, and the local planning algorithm converts the fold trajectory to a series of arc trajectories, taking care to avoid the obstacles not effectively recognized by the global planning, and optimizes the local trajectories according to certain guidelines.

During the global path planning, the mast movement is used to drive the laser rangefinder to scan and image the terrain around the lunar rover, and then the navigation control unit of the lunar rover completes the 3D terrain stitching, and generates the sequential motion target points. These points are submitted to the local obstacle avoidance planning algorithm, which completes the fusion with the large 3D terrain data recovered from the camera, and plans the safest path to the nearest target point. The path is handed over to motion control for motion decomposition to start moving. After reaching the target point, the lunar rover automatically updates the next target point, and continues to move until it finishes all the path points and reaches the final target.

## 3.6 Data communication

As discussed earlier, Optical fiber is used for communication between the master station and other stations. All the scientific data are collected at the master station of the array. Since the array is deployed on the far side of the Moon, the master station need to transmit the scientific data back to the Earth through a relay communication satellite of the Moon. The orbiter of the LARAF, or a successor of the Queqiao (Magpie Bridge) relay communication satellite is needed to fulfill this role.

If an orbiter is used for relay communication, it could be one run in a 16000km×500km elliptical orbit around the Moon, using a Ka-band relay antenna with 1.6 m aperture. The master station upload scientific data during the lunar day and the uplink rate is not less than 15Mbps. Over 1500GB scientific data can be uploaded in one lunar day.

## 3.7    Site determination

The main requirement of the site for our array is to have a relatively flat region for easy landing and deployment. The array should not be too close to the regions near the near-far side edge, to avoid being interfered by the diffracted ultralong wave from the Earth. A mid-latitude site is desirable, as it allows viewing a large part of sky than the polar region, and is convenient for solar energy use.

We try to find a suitable terrain between 20°S~50°S and 120°W~120°E on the lunar farside. This area crosses the northern side of the Aitken Basin and connects some highlands at the edge of the basin (Fig.14). The Chang'e-4 lunar probe landed in this area. Based on the data from Chang'E-1/2 (China), the Lunar Reconnaissance Orbiter (USA, LRO), and SELENE(Japan), the terrain of the lunar middle and low latitude regions is mostly large lunar mare, with some independently distributed mountains. The elevation variations are small for most of the region. The terrain at the junction of the Aitken Basin and nearby mountains is slightly undulating, but the interior of the Aitken Basin is still dominated by a large area of flat lowlands, with an elevation difference of about 19 km in the lunar mid-low latitude region. In order to meet the requirement of soft landing on the lunar surface and deployment of LARAF, we screened the area outside impact craters with a slope of <8° in patches (>10km×10km) as potential landing points, and found a number of candidate sites.

In the end, after comparing various conditions, we selected a tentative candidate site at latitude -171.4947° and longitude -32.7393°. The average slope of this area is about 7°, with 62.3% of the area having slope <8°. We shall choose one that is far from the mountain rims.

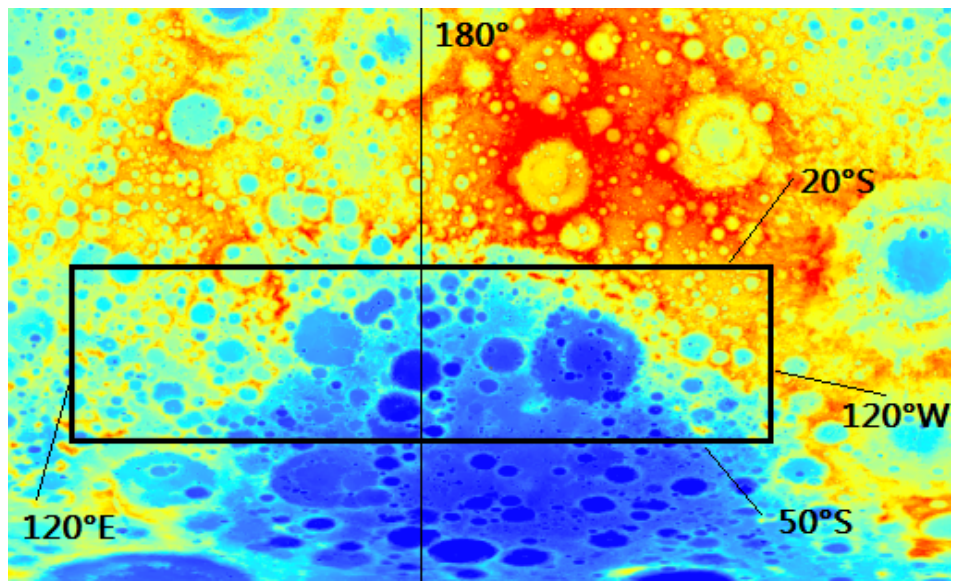

Figure 14.      Deployment area

# 4  The Mission Plan

## 4.1   System Composition

LARAF is planned to be launched by a Long March 5 Series Launch Vehicle, and the probe consists of a propulsion module, a lander (including the lunar rover), and a relay satellite (can be omitted if there is already relay satellite in operation). The propulsion module is responsible for completing the Earth-Moon transfer of the assembly. The lander fully inherits the lunar soft landing technology of Chang'e-5, including the lunar landing, carrying scientific payloads, lunar rover and other equipment. After landing, the lander will also serve as the master station of the LARAF. The intelligent lunar rover carried on the lander is responsible for the autonomous deployment of the stations and connections. The Relay Satellite is released during the lunar orbit and is responsible for the data transmission relay mission during LARAF operation. The effect of the LARAF deployment is shown in fig. 15, though in reality the Earth would not be visible from the far side.

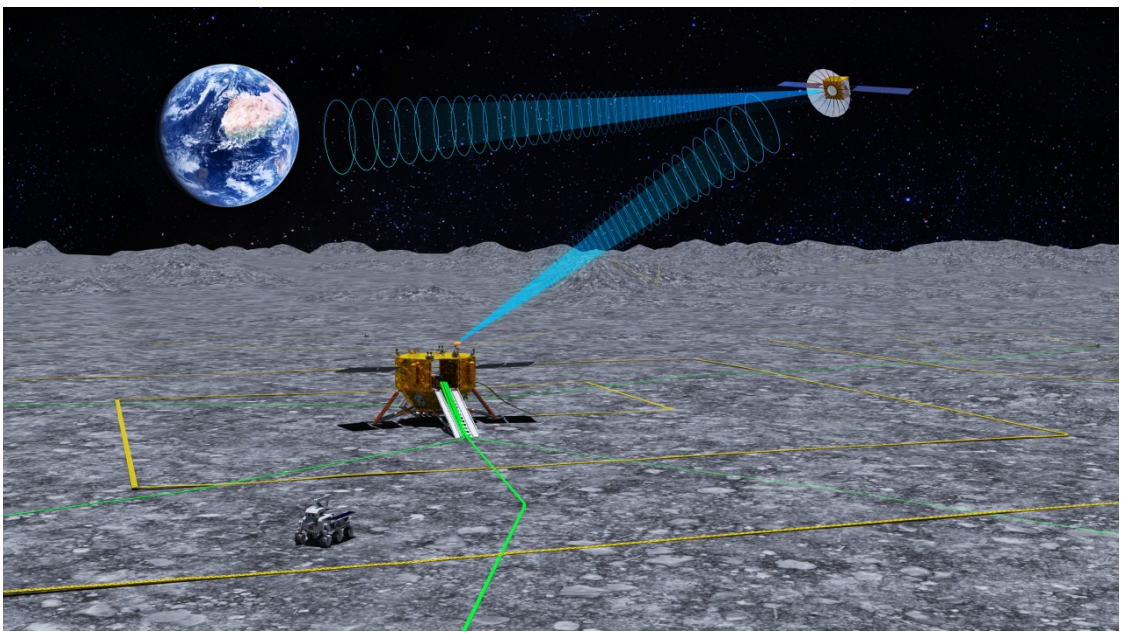

Figure 15.        The working effect of the lunar dorsal low-frequency radio detection array

## 4.2　Overall Indicators and Mission Process

The overall indicators for the mission is given in Table.2.

The combination will be launched into the Earth-Moon transfer orbit, with perigee of 200km and apogee of 380,000km, and then transferred to the large elliptical orbit around the Moon. After the Relay Satellite is released, the Lander is separated from the propulsion module. The Lander flies around the Moon and descends to a 200km circular orbit, then starts a powered descent, and lands on the Moon far-side. The process is shown in Figure 16.

It will take 8 months to deploy the antenna array using the rover. The lunar surface initialization is completed during the lunar day, including the deployment of master station antenna, station, station antenna. On the lunar night, the Lunar Rover returns to the master station and enters the sleep mode using the power supply of the master station.

After the deployment is successfully completed, science operation starts. The astronomical observation will be carried out mostly during lunar night, while during the lunar day the instruments are mostly powered off. However, at the central station the RFC will be charged at this time, and the data is downlinked to Earth via the relay satellite, making it ready for taking observation at the next lunar night.

Table 1.  Overall Indicators

| Serial number | Projects | Indicators |
|---|---|---|
| 1 | Radio Array Deployment Locations | Lunar farside 20°~50°N latitude |
| 2 | Orbiter track | 16000km×500km frozen track |
| 3 | Lander Lunar Night Power Consumption | ≤620W |
| 4 | Radio array supply voltage | 1000V |
| 5 | Radio array deployment time | 8 Months |
| 6 | Radio array sensitivity (SEFD) | 20000Jy |
| 7 | Spatial resolution of radio arrays | 20′@10MHz |
| 8 | Radio array field of view | >80°×80° |

| 9  | Radio array operating time | Monthly Night |
|----|----------------------------|---------------|
| 10 | Load working temperature   | -100℃~50℃    |
| 11 | Lifespan                   | 5 years       |

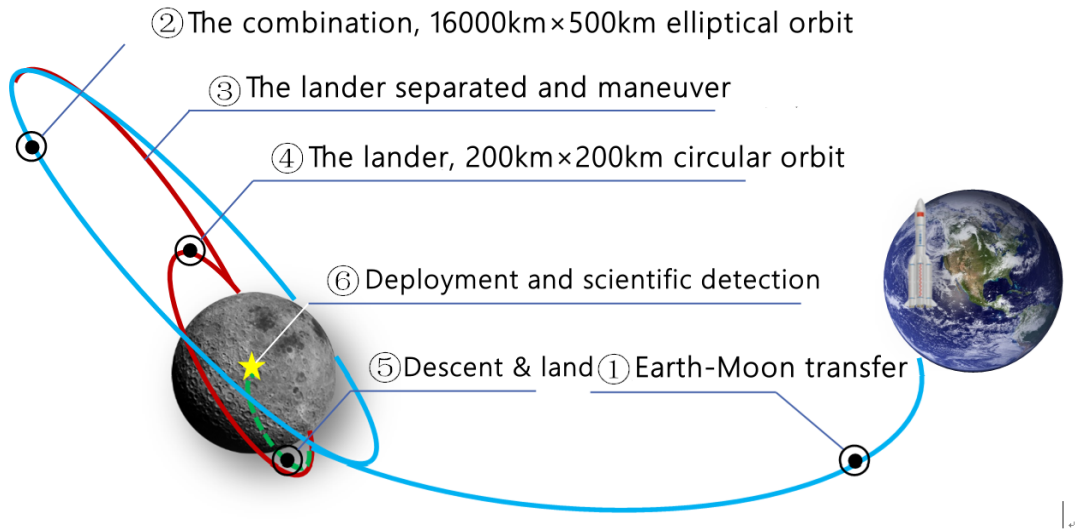

Figure 16.      Flight process

# 5  Conclusion

We have considered the concept of a low frequency radio interferometer array on the far side of the Moon. The array consists of a master station and 20 stations, arranged in a spiral configuration. Each station contains 12 antennas, printed on a long membrane. The membrane also contain optical fibers and power wire to the station. At the station the signal is digitized and transmitted to the master station.

We considered various options for the power supply, data communication, deployment, and position determination. We find that in near terms, the RFC is probably the most plausible power supply. To take advantage of the better observing condition at lunar night for most science cases, we propose to carry out the observation in the night, while charging and sleep during lunar day. The stations will be deployed by a lunar rover, which will carry the stations to the designated positions during the lunar day, and return to the lander for the night. The deployment will be completed in 8 months. The rover will make automatic path planning for its movement, and also determine the relative position for each station. The whole system can be launched with the Long March 5 rocket.

The sensitivity of this array is sufficiently high, so that it has the potential of detecting many interesting objects, such as radio galaxies and quasars, supernovae remnants, exoplanets, etc, and produce much needed data for ultralong wavelength astronomy. The technology developed and tested in this mission will also form an indispensable foundation for the future huge arrays designed for measuring primordial fluctuations of the dark ages.

## Acknowledgments
This research is supported by the National Key R&D Program of China (Grant No. 2022YFF0504300) and the Chinese Academy of Science (Grant No. ZDKYYQ20200008). XC acknowledge the Royal Society for providing travel support for attending the Royal Society Meeting in London.

**Data Accessibility**
This article has no data.

**Authors' Contributions**

**Competing Interests**
We have no competing interests.